\documentclass[twocolumn,letterpaper,showpacs,aps,prl,superscriptaddress,preprintnumbers,amsmath,amssymb]{revtex4}
\usepackage{graphicx}
\usepackage{dcolumn}
\usepackage{bm}

\newcommand{\BABARPubYear}    {06}
\newcommand{\BABARPubNumber}  {019}
\newcommand{\SLACPubNumber} {11786}

\usepackage{relsize}
\usepackage{xspace}
\def\babar{\mbox{\slshape B\kern-0.1em{\smaller A}\kern-0.1em
    B\kern-0.1em{\smaller A\kern-0.2em R}}}
\def\pep2{PEP-II}
\def\D0bar{\kern 0.2em\overline{\kern -0.2em D}{\kern 0.1em}\xspace^0}

\newcommand{\kev}{\ensuremath{\mathrm{\,ke\kern -0.1em V}}\xspace}
\newcommand{\gev}{\ensuremath{\mathrm{\,Ge\kern -0.1em V}}\xspace}
\newcommand{\mev}{\ensuremath{\mathrm{\,Me\kern -0.1em V}}\xspace}
\newcommand{\gevc}{\ensuremath{{\mathrm{\,Ge\kern -0.1em V\!/}c}}\xspace}
\newcommand{\mevc}{\ensuremath{{\mathrm{\,Me\kern -0.1em V\!/}c}}\xspace}
\newcommand{\gevcc}{\ensuremath{{\mathrm{\,Ge\kern -0.1em V\!/}c^2}}\xspace}
\newcommand{\mevcc}{\ensuremath{{\mathrm{\,Me\kern -0.1em V\!/}c^2}}\xspace}
\newcommand{\kevcc}{\ensuremath{{\mathrm{\,ke\kern -0.1em V\!/}c^2}}\xspace}

\def\piz   {\ensuremath{\pi^0}\xspace}
\def\pip   {\ensuremath{\pi^+}\xspace}

\def\Kbar{\kern 0.2em\overline{\kern -0.2em K}{}\xspace}

\def\Kpi{K^-\pi^+}
\def\Kpipi{K^-\pi^+\pi^+}
\def\Ktpi{K^-\pi^+\pi^-\pi^+}

\def\LamTE{\Lambda_c(2880)^+}
\def\LamTN{\Lambda_c(2940)^+}

\begin{document}

\preprint{BABAR-PUB-\BABARPubYear/\BABARPubNumber}
\preprint{SLAC-PUB-\SLACPubNumber}

\title{\boldmath Observation of a Charmed Baryon Decaying to $D^0 p$
at a Mass Near 2.94~\gevcc}

%
\author{B.~Aubert}
\author{R.~Barate}
\author{M.~Bona}
\author{D.~Boutigny}
\author{F.~Couderc}
\author{Y.~Karyotakis}
\author{J.~P.~Lees}
\author{V.~Poireau}
\author{V.~Tisserand}
\author{A.~Zghiche}
\affiliation{Laboratoire de Physique des Particules, F-74941 Annecy-le-Vieux, France }
% ...
\collaboration{The \babar\ Collaboration}
\noaffiliation

\date{\today}

\begin{abstract}
A search for charmed baryons decaying to $D^0 p$
reveals two states: the $\LamTE$ baryon and a previously unobserved
state at a mass 
of [$2939.8\pm 1.3\;\text{(stat.)}\pm 1.0\;\text{(syst.)}$]~\mevcc and
with an intrinsic width of 
[$17.5\pm 5.2\;\text{(stat.)}\pm 5.9\;\text{(syst.)}$]~\mev.
Consistent and significant signals
are observed for the $\Kpi$ and $\Ktpi$ decay modes of the $D^0$ in 287 
${\rm fb}^{-1}$  annihilation
data recorded by the 
\babar\  detector at a center-of-mass energy of 10.58~\gev. 
There is no evidence in the $D^+ p$ spectrum
of doubly-charged partners. The mass and intrinsic width
of the $\LamTE$ baryon and relative 
yield of the two baryons are also measured.
\end{abstract}

\maketitle

Charmed baryons are expected to exhibit a rich spectrum 
of states. Only a few of these states have been 
confirmed~\cite{Eidelman:2004wy}. The heaviest singly-charmed baryon
previously observed is the $\LamTE$
decaying to $\Lambda_c \pi^+\pi^-$~\cite{Artuso:2000xy}.
The $\LamTE$ baryon is notable not only due to its
narrow width ($<8$~\mev) but also because it one of only two
singly-charmed bayrons, along with the
$\Xi_c(2815)$~\cite{Alexander:1999ud}, 
found above the $Dp$ mass threshold.

Presented in this Letter is the observation of a new charmed
baryon decaying to $D^0 p$~\cite{cpfootnote} 
with a mass of approximately 2.94~\gevcc\  and
an intrinsic width of approximately 20~\mev. This baryon, tentatively
labeled the $\LamTN$, is
observed in 287~${\rm fb}^{-1}$ of $e^+e^-$ annihilation
data collected near $\sqrt{s} = 10.58$~\gev
by the \babar\ detector~\cite{Aubert:2001tu} at the \pep2
asymmetric-energy $e^+e^-$ storage rings. Along with this new baryon,
the decay $\LamTE \to D^0 p$ is also observed.
The masses, intrinsic widths of both baryons and their relative production
rate are measured.

The goal of this analysis is to study the inclusive $D^0 p$ mass
spectrum. Two samples of $D^0$ mesons are identified using the $\Kpi$
and $\Ktpi$ final states. Each sample is produced by
combining charged tracks of the appropriate composition in a geometric
fit to a common vertex. The $\chi^2$ probability of this fit is
required to exceed 2\%.
Charged particle species ($K^+,\pi^+,p$) are separated
using a likelihood algorithm that combines data from
a ring-imaging Cherenkov detector
with the measured energy loss in the tracking 
systems~\cite{Aubert:2001tu}.
Each proton candidate is combined with each $D^0$
candidate using a geometric vertex
fit that assumes a common production point within the nominal beam envelope. 
The $\chi^2$ probability of this fit is
required to be better than 2\%.

Requirements are imposed on three additional quantities to improve
the signal purity of the $D^0 p$ samples:
$\Delta m$, the difference 
between the reconstructed
$D^0$ mass and the accepted value of 
$m_{D^0} = 1864.6$~\mevcc~\cite{Eidelman:2004wy}; $p^*$,
the center-of-mass momentum of the $D^0 p$ system;
and $\cos\vartheta$,
where $\vartheta$ is angle of the proton with respect to the $e^+e^-$
system in the $D^0 p$ center-of-mass frame.
For isotropic production (expected for the $\LamTN$), the $\cos\vartheta$
distribution will be flat whereas background tends to peak at $\pm 1$.
Studies of Monte Carlo (MC) simulated data samples are used to
determine the specific requirements on these quantities that maximize
the expected significance of signals introduced in the mass
region near 2940~\mevcc.
The resulting best criteria 
are $|\Delta m| < 14$~\mevcc, $p^* > 2.6$~\gevc, and
$\cos\vartheta < 0.8$ for the $D^0 \to \Kpi$ sample and
$|\Delta m| < 9$~\mevcc, $p^* > 2.8$~\gevc, and
$\cos\vartheta < 0.8$ for the $D^0 \to \Ktpi$ sample.
The $\Delta m$ requirements correspond to approximately
two standard deviations in $D^0$ mass resolution.
The $p^*$ requirement removes all sources of $D^0 p$
combinations from $B$ meson decay.

A MC simulation of a baryon of mass
2.94~\gevcc\  decaying to $D^0 p$ predicts selection efficiencies
between 30\% and 38\% for the $D^0 \to \Kpi$ final state
depending on $p^*$
and between 12\% and 14\%  for the $D^0 \to \Ktpi$
final state. A proton purity
of approximately 83\% in the final $D^0 p$ sample
is estimated from studies of a comparable MC sample.

\begin{figure}
\begin{center}
\includegraphics[width=\linewidth]{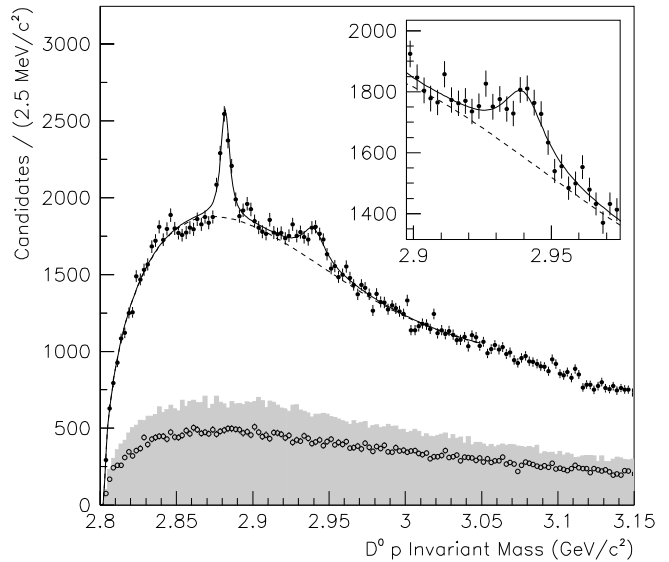}
\caption{\label{fg:fit.both}The solid points are
the $D^0 p$ invariant mass distribution
of the final sample. Also shown are (gray) the contribution from
false $D^0$ candidates estimated from $D^0$ mass sidebands and
(open points) the mass distribution from
wrong-sign $\overline{D}{}^0 p$ candidates.
The solid curve is the fit described in the text.
The dashed curve is the portion of that fit attributed to combinatorial
background.}
\end{center}
\end{figure}

To calculate a $D^0 p$ invariant mass, each $D^0$
candidate is assigned an energy that is consistent with a $D^0$
mass of $m_{D^0}$. The resulting combined $D^0 p$ invariant mass 
spectrum is shown in Fig.~\ref{fg:fit.both}. Two peaks are apparent.
The clear signal at 2.88~\gevcc\  is likely due to the decay of the
$\LamTE$ baryon. The signal at 2.94~\gevcc\  is the evidence for
the new $\LamTN$ baryon. 
No similar structures are observed in the wrong-sign $\overline{D}{}^0 p$
candidate combinations. Candidates selected from $D^0$ mass
sidebands are used to estimate the contribution 
from non-$D^0$ sources (see Fig.~\ref{fg:fit.both}). 
This sideband sample shows no structure.

An unbinned likelihood fit is used to model the $D^0 p$ spectrum
from the kinematic limit up to 3.05~\gevcc.
This fit includes $\LamTE$ and $\LamTN$ states, each modeled by
a relativistic Breit-Wigner lineshape $\sigma(m)$
convolved with a Gaussian resolution function.
The Breit-Wigner line shape $\sigma(m)$ is:
\begin{equation}
\sigma(m) \propto \frac{ q(m) }
{ \left( m^2 - m_0^2 \right)^2 + m_0^2 \Gamma^2 } \;,
\label{eq:bw}
\end{equation}
where $\Gamma$ is the intrinsic width and is constant (i.e. not mass
dependent), $m_0$ is the mass pole, and $q$ is the three-momentum
magnitude of the $D^0$ or proton in the $D^0 p$ rest frame for 
a given mass $m$. The detector resolution is obtained from
MC simulation which predicts 1.8~\mevcc\  and
1.3~\mevcc\   for the $D^0 \to \Kpi$ and $D^0 \to \Ktpi$ samples,
respectively. 

The product of a fourth-order polynomial and two-body phase 
space~\cite{Eidelman:2004wy}
is used to model the combinatorial background. A fit based on
this background shape and the $\LamTE$ and $\LamTN$ signals
is shown in Fig.~\ref{fg:fit.both} and
results in a $\LamTN$ mass of $2939.8\pm 1.3$~\mevcc, a width
of $17.5\pm 5.2$~\mev, and a raw yield of 
$2280\pm 310$ decays (statistical errors only). The $\LamTE$
properties obtained are a mass of $2881.9\pm 0.1$~\mevcc\  and
a width of $5.8\pm 1.5$~\mev, consistent with the
CLEO results~\cite{Artuso:2000xy}, and a raw yield
of $2800 \pm 190$ decays (statistical errors only).
If the $\LamTN$ signal is removed from the fit,
the log likelihood changes by 38.2, which
is equivalent (in one degree of freedom) to a signal significance
of 8.7 standard deviations. 
If the $D^0 \to \Kpi$ and $D^0 \to \Ktpi$
samples are fit separately, the resulting masses, widths, and relative
yields of the $\LamTE$ and $\LamTN$ baryons are consistent within
statistical errors. After accounting for selection efficiency
and $D^0$ branching fractions, the absolute yields for the two
$D^0$ decays modes are consistent for both the
$\LamTE$ and $\LamTN$ baryons.

\begin{figure}
\begin{center}
\includegraphics[width=0.857\linewidth]{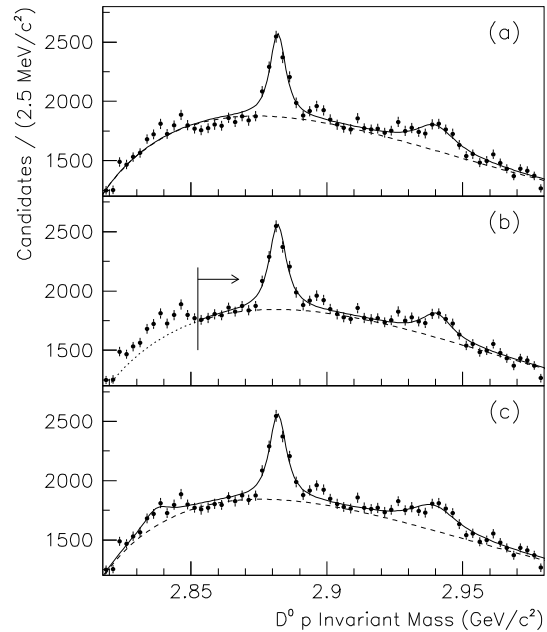}
\caption{\label{fg:fit.examplesyst}Three examples of how the
structure near a $D^0 p$ mass of 2.84~\gevcc\  can be modeled.
Shown are the results of fits that
(a) assume a smooth distribution (as used for the central result) 
(b) exclude data below a
mass of 2.8525~\gevcc, and (c) add an extra resonance contribution.}
\end{center}
\end{figure}

The above likelihood fit models the mass spectrum near 2.84~\gevcc\  
as a smooth distribution (Fig.~\ref{fg:fit.examplesyst}(a)). 
There is, however, a non-distinct structure
near a mass of 2.84~\gevcc\  whose origin is not understood,
and so this model may not be accurate. Various modifications of the
fit are employed as systematic checks. 
At one extreme, if the
likelihood fit is limited to masses above 2.8525~\gevcc\   
(Fig.~\ref{fg:fit.examplesyst}(b)), the
result is a substantial decrease (29\%) in the $\LamTN$ yield,
a 0.5~\mevcc\  shift in mass, and a smaller width (12.5~\mev). 
The changes in the fitted $\LamTN$
properties are much smaller if a third signal line shape (of variable
mass and width) is added
to the fit (Fig.~\ref{fg:fit.examplesyst}(c)). 
None of these alternate fits lead to a reduction in
the statistical significance
of the $\LamTN$ signal below 7.2 standard deviations.

Because the $\LamTE$ and $\LamTN$ are near the $D^0 p$ threshold,
the systematic uncertainty in mass
from possible detector biases is relatively
small. This uncertainty is calculated by considering appropriate variations
in the assumed $B$ field strength and detector material using a procedure
developed for measuring the $\Lambda_c$ mass~\cite{Aubert:2005gt}. This 
procedure is also
used to calculate small ($<0.1$~\mevcc) corrections to the 
reconstructed $D^0 p$ mass.
An additional uncertainty of 0.5~\mevcc\  arises 
from the current knowledge of $m_{D^0}$. The results for
the $\LamTN$ baryon are:
\begin{equation*}
\renewcommand{\arraystretch}{1.2}
\begin{array}{r@{\;}c@{\;}r@{\;}c@{\;}r@{\;}l@{\;}c@{\;}r@{\;}l@{\;}l}
m &=[& 
      2939.8 &\pm& 1.3 &\text{(stat.)}& \pm& 1.0 &\text{(syst.)}& ]~\mevcc \\
\Gamma &=[& 
        17.5 &\pm& 5.2 &\text{(stat.)}& \pm& 5.9 &\text{(syst.)}& ]~\mev\;.
\end{array}
\end{equation*}
For the $\LamTE$ baryon the results are:
\begin{equation*}
\renewcommand{\arraystretch}{1.2}
\begin{array}{r@{\;}c@{\;}r@{\;}c@{\;}r@{\;}l@{\;}c@{\;}r@{\;}l@{\;}l}
m &=[& 
      2881.9 &\pm& 0.1 &\text{(stat.)}& \pm& 0.5 &\text{(syst.)}& ]~\mevcc \\
\Gamma &=[& 
         5.8 &\pm& 1.5 &\text{(stat.)}& \pm& 1.1 &\text{(syst.)}& ]~\mev\;.
\end{array}
\end{equation*}
From the baryon yields obtained from the likelihood fits, the following
ratio of production cross sections and decay branching ratios is calculated:
\begin{align*}
\frac{\sigma(\LamTN)\mathcal{B}r(\LamTN\to D^0 p)}
     {\sigma(\LamTE)\mathcal{B}r(\LamTE\to D^0 p)} \nonumber \\
     \quad\quad = 0.81 \pm 0.13 \;\text{(stat.)} \pm 0.35\;\text{(syst.)} \;,
\end{align*}
where the systematic uncertainty is dominated by uncertainties in
the background shape.

Various tests are applied to the data to confirm the $\LamTN$ signal.
Since the signal is observed in two different $D^0$ decay modes, it appears
to be associated with real $D^0$ decays.
The lack of any structure in the $D^0$ sideband samples and the 
relative size of these samples support this conclusion. Since the
sample of protons is 83\% pure, it is unlikely that the
$\LamTN$ signal could arise from proton mis-identification.
As further confirmation, 
when the $K^+$ or $\pi^+$ mass is assigned 
to the protons, the resulting $D^0 K^+$ and $D^0 \pi^+$ invariant mass
distributions show no evidence of structure.

Even if the observed signal is attributed to a combination of $D^0$
and protons, it is still possible to produce a false signal 
from the reflection of heavier
states. 
One example of such
a possible reflection is a hypothetical 
baryon of mass near 3.10~\gevcc\  decaying
to either $D^*(2010)^+ p$ or $D^*(2007)^0 p$. Such a baryon, if 
sufficiently narrow,  would produce a $D^0 p$ mass spectrum
(after ignoring the $\pip$ or $\piz$ from $D^*$ decay) 
of approximately the correct mass and width. Such a baryon would also
be clearly visible in the $D^*(2010)^+ p$ or $D^*(2007)^0 p$ mass
distributions. An explicit search in those mass distributions shows no signal,
and thus this hypothesis is strongly disfavored.

Another possible reflection is from a baryon of mass 3.13~\gevcc\  
decaying to $D^0 \Sigma^+$. The kinematics of such a decay could
produce peaks at both 2.85~\gevcc\   and 2.94~\gevcc\  if the
$\Sigma^+$ had the appropriate spin alignment. The
$\Sigma^+$, however, is a long-lived particle, and MC studies indicate
that for this decay the proton vertex $\chi^2$ probability distribution 
would peak at zero.
An investigation of the $\chi^2$ probability of the 
$\LamTN$ signal seen in the data indicates a flat distribution.
Thus, a reflection from $D^0 \Sigma^+$ decay is also strongly disfavored.

The simplest interpretation of the $\LamTN$ signal is that it arises
from a charmed
baryon of quark content $cdu$. Under this scenario
the decay to $D^0 p$ involves
simple $u\overline{u}$ gluon splitting.
The remaining question is whether the $\LamTN$ belongs to an isotriplet.
The most direct way to address this question is to explicitly search
for a neutral or doubly-charged partner of nearly the same mass and
width, analogous to the $\Sigma_c^0$ and $\Sigma_c^{++}$.
The \babar\  detector cannot isolate the most
obvious neutral decay mode ($D^0 n$). It is possible, however,
to search for a doubly-charged baryon decaying to $D^+ p$.

To select a sample of $D^+$ candidates, the same methods
used for the $D^0$ samples are applied to the decay $D^+ \to \Kpipi$.
The selection requirements for the $D^+ p$ sample are
$|\Delta m| < 12$~\mevcc, $p^* > 2.7$~\gevc, and $\cos\vartheta < 0.8$.
The efficiency for this selection is approximately 23\%.

\begin{figure}
\begin{center}
\includegraphics[width=0.714\linewidth]{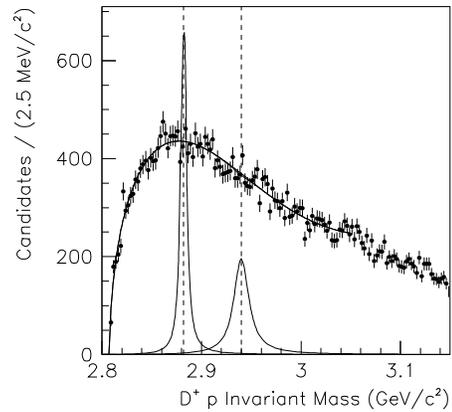}
\caption{\label{fg:fit.dplus}The invariant mass distribution
of selected $D^+ p$ candidates. The curve 
is the result of the fit described in the text. The curves
below are the lineshapes of 
the $\LamTE$ and $\LamTN$ baryons obtained from the $D^0 p$ data,
drawn approximately to scale after correcting for selection efficiency
and $D^0$ and $D^+$ branching fractions.}
\end{center}
\end{figure}

The resulting $D^+ p$ distribution is shown in Fig.~\ref{fg:fit.dplus}.
No signals corresponding to either
the $\LamTE$ or $\LamTN$ baryon are apparent.
A likelihood fit which
assumes a doubly-charged partner of the $\LamTN$ of identical mass
and width results in a yield of $-40 \pm 120$ candidates
(statistical error only).

Based on previous observations, such as the
CLEO measurement of the $\Sigma_c^0$ and $\Sigma_c^{++}$~\cite{Artuso:2001us},
one would expect similar production rates
for the $\LamTN$ and a hypothetical doubly-charged partner. Under 
the additional assumption that the branching fraction of the
doubly-charged baryon to $Dp$ is the same,
the expected doubly-charged
signal yield would be approximately $2200$ decays
once the $D^0$ and $D^+$ branching fractions and 
selection efficiencies are accounted for (see Fig.~\ref{fg:fit.dplus}). 
It thus seems unlikely
that a doubly-charged partner exists, unless its production
is largely suppressed or it decays in an unexpected fashion.

The $\LamTN$ baryon is interesting for several reasons.
Relativistic quark model calculations~\cite{Migura:2006ep}
predict three excited
$\Lambda_c$ baryons of different spin-parity quantum numbers
near a mass of 2.94~\gevcc.
The $DN$ decay mode, although not 
unexpected~\cite{Pirjol:1997nh,Blechman:2003mq},
is a final state that has received relatively little theoretical 
investigation.
If this baryon had a significant branching fraction
to $\Lambda_c \pi^+\pi^-$ it probably would have been observed
with the $\LamTE$ by CLEO~\cite{Artuso:2000xy}. It is not clear,
however, why this particular decay mode, which is favored by 
phase space, is suppressed.
One observation which is notable, even if it might be a
simple coincidence, is that at a mass of 2939.8~\mevcc, the $\LamTN$
is just 6~\mevcc\  below the $D^{*0} p$ threshold. It is also 
interesting that the $\LamTN$ is approximately one pion mass
heavier than the $\Sigma_c(2800)^+$, a charmed baryon recently 
discovered by BELLE~\cite{Mizuk:2004yu} decaying to $\Lambda_c \pi^0$.

The $\LamTE$ mass and width results presented here are 
consistent with but more precise than
the CLEO measurement of $m = 2880.9 \pm 2.3$~\mevcc\  and 
$\Gamma < 8$~\mev\ (at 90\% CL).
The existence of the decay $\LamTE \to D^0 p$ rules out
various interpretations of this baryon~\cite{Blechman:2003mq}.

\begin{acknowledgments}
We are grateful for the excellent luminosity and machine conditions
provided by our \pep2\ colleagues, 
and for the substantial dedicated effort from
the computing organizations that support \babar.
The collaborating institutions wish to thank 
SLAC for its support and kind hospitality. 
This work is supported by
DOE
and NSF (USA),
NSERC (Canada),
IHEP (China),
CEA and
CNRS-IN2P3
(France),
BMBF and DFG
(Germany),
INFN (Italy),
FOM (The Netherlands),
NFR (Norway),
MIST (Russia), and
PPARC (United Kingdom). 
Individuals have received support from CONACyT (Mexico), 
Marie Curie EIF (European Union),
the A.~P.~Sloan Foundation, 
the Research Corporation,
and the Alexander von Humboldt Foundation.

\end{acknowledgments}

\bibliography{note1408}

\begin{thebibliography}{11}
\bibitem[{cpf()}]{cpfootnote}
\bibinfo{note}{Inclusion of charge conjugate states is implied throughout this
  paper.}
\end{thebibliography}

\end{document}